\documentclass{phostproc}

\title{R mode oscillations ubiquitous in stars}
\author{Hideyuki Saio 
        }

\affiliation{Astronomical Institute, Graduate School of Science, Tohoku University, Sendai, Japan}

\shorttitle{R mode oscillations}
\shortauthors{H. Saio}

\abs{
We discuss properties and examples of r-mode (large scale Rossby wave) oscillations in rotating stars. Motions are nearly toroidal but affected weakly by buoyancy. R modes seem to be excited mechanically in most cases, and  ubiquitous in various types of rotating stars;e.g., fast rotating $\gamma$ Dor stars, spotted stars, eclipsing binary stars, Be stars, and accreting white dwarfs. In the Fourier spectrum of a star, even (i.e., symmetric to the equator) r modes of azimuthal order $m$ appear as a group of frequencies just below $m$ times the rotation frequency. From this property, a rotation frequency can be determined if r modes are detected. We discuss r modes in KIC~8462852 (Boyajian's star), KIC~9117875 (Am star), KIC~6128830 (HgMn star), and the accreting white dwarf in the dwarf nova GW~Lib.
}

\begin{document}

\maketitle

\section{Introduction}
R-mode oscillations are normal modes of global Rossby waves. The restoring force arises from the conservation of the vertical component of total vorticity under the presence of the latitudinal dependence of $\Omega_r(=\Omega\cos\theta)$ \citep[see e.g][for discussion]{sai82}, where $\mathbf{\Omega}$ is the angular velocity of rotation.
Although toroidal motions dominate, small radial motions interact with buoyancy so that r modes have dense frequency spectra similarly to the g modes.
  
The possibility of r mode oscillations in rotating stars has been discussed theoretically for long time \citep[e.g.,][]{pap78,pro81,sai82,ber83,dzi87}. 
But the first clear signature of r modes has been found only after the extremely precise light curves by the Kepler satellite became available; \citet{vanr16} found for the first time a clear r-mode signature in period-spacing/period relations of several fast rotating $\gamma$ Dor stars.
Considering visibility distributions of r modes under the assumption of energy equipartition, \citet{sai18} found another r-mode signature, a dense frequency group to the left of the rotation frequency in Fourier amplitude spectra ("hump\&spike").
The visibility distribution is found to be consistent with the amplitude distribution of r modes in $\gamma$ Dor stars found by \citet{vanr16}.  
We identified the r-mode signature with the 'RotD' feature that was found in many early-type main sequence stars by e.g. \citet{bal13,bal15,bal17}. 
\citet{sai18} also found the r-mode signature in eclipsing binary stars and Be stars.

In this paper, after a brief summary on the property of r and g modes in a rotating star, we extend our search for the r-mode signature to other types of stars; Boyajian's star, Am and HgMn stars, and accreting white dwarfs in dwarf novae.

\section{Low-frequency oscillations in a rotating star}
Under the Cowling approximation, where the Eulerian perturbation of gravitational potential is neglected, basic equations for adiabatic nonradial pulsations in a rotating star in the co-rotating frame are given as \citep[see e.g.][]{unno}
\begin{equation}
{\partial\mathbf{u}\over\partial t} = -{1\over\rho}\nabla p'- g{\rho'\over\rho}{\bf e}_r -2\mathbf{\Omega}\times\mathbf{u}
\quad \& \quad
{\partial\rho'\over\partial t} = -\nabla\cdot(\rho\mathbf{u})
\label{eq:gov}
\end{equation}
with the adiabatic condition
\begin{equation}
p'-g\rho\xi_r= c_{\rm s}^2\left(\rho'+\xi_r{d\rho\over dr}\right),
\end{equation}
where $~'~$ means Eulerian perturbation, $g$ local gravitational acceleration, $c_{\rm s}$ adiabatic sound speed. 
%$\mathbf{\Omega}$ angular frequency of rotation. 
Pulsation velocity vector $\mathbf{u}$ in the co-rotating frame is given as, $\mathbf{u} = i\omega\vec{\xi}$ with $\omega$ and $\vec{\xi}$ being pulsational (angular) frequency and displacement in the co-rotating frame, respectively.
The term $2\mathbf{\Omega}\times\mathbf{u}$ is Coriolis force, while rotational deformation of the equilibrium structure is neglected.

For low-frequency oscillations, in which horizontal motion is much larger than vertical motion, traditional approximation of rotation (TAR) is very useful. 
In the TAR, the horizontal component of angular velocity of rotation, $\Omega\sin\theta$ is neglected in the governing equations given in equation (\ref{eq:gov}). 
Then, the angular dependence of pulsation variables is separated from the radial dependence as
\begin{equation}
(\xi_r, p', \rho') = (\xi_r(r), p'(r), \rho'(r)) \Theta_k^m(\theta; s){\rm e}^{i\omega t + im\phi},
\label{eq:func}
\end{equation}
and the governing equations are reduced to 
\begin{equation}
-\omega^2\rho\xi_r =-{dp'\over dr}- g\rho' 
\label{eq:tar1}
\end{equation}
and
\begin{equation}
\rho'=-{d(r^2\rho\xi_r)\over r^2d r}+{\lambda(s,k)\over\omega^2r^2}p',
\label{eq:tar2}
\end{equation}
where $\Theta_k^m$ and $\lambda$ are eigenfunction and eigenvalue of Laplace's tidal equation \citep[see e.g.,][]{lee97,tow03a}.
Both $\Theta_k^m$ and $\lambda$ depend on the spin parameter
\begin{equation}
s \equiv {2\Omega\over\omega},
\end{equation}
and parameter $k$ is introduced for ordering $\lambda$ and $\Theta_k^m$.
$\Theta_k^m$ is symmetric with respect to the equator if $|k|$ is a even number or zero, while it is antisymmetric if $|k|$ is a odd number.  

\begin{figure}
  \centering
  \includegraphics[width=\columnwidth]{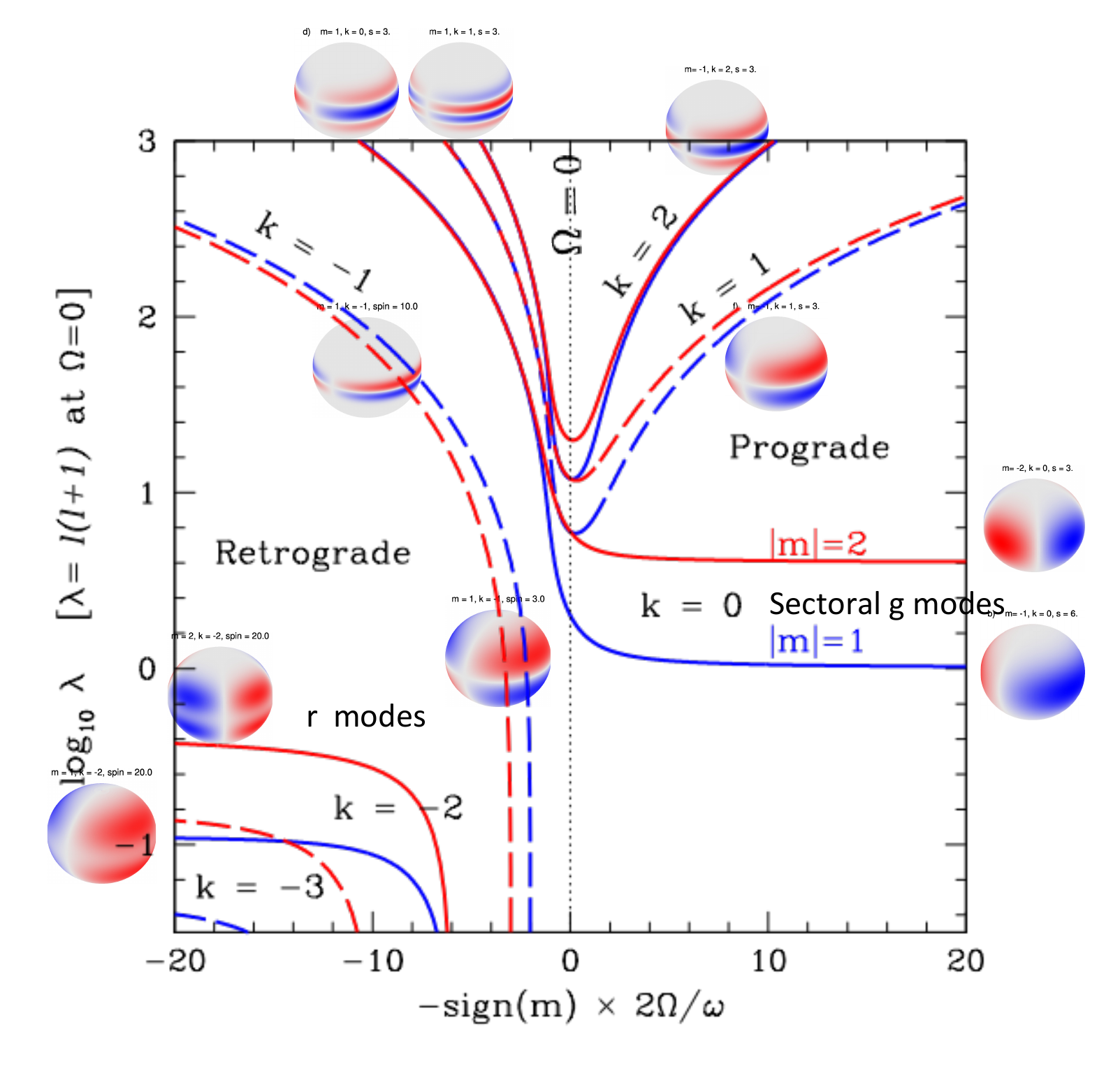} %{spin_lambda.pdf}
  \caption{Values of $\lambda$ (= eigenvalue of Laplace tidal equation) are plotted with respect to spin parameter, $2\Omega/\omega$ with $\Omega$ and $\omega$ being rotational and pulsational (in the co-rotating frame) angular frequencies, respectively. In this paper we adopt the (classical) convention that $m<0$ for prograde modes and $m>0$ for retrograde modes. The left and right sides correspond retrograde and prograde modes in the co-rotating frame, respectively. Blue and red lines are for modes of $|m|=1$ and $2$, respectively. For a larger  value of $\lambda$ amplitude is more strongly confined to equatorial zone as shown by pulsation patterns at some places. }
\label{fig:lambda}
\end{figure}

Figure~\ref{fig:lambda} shows the variation of $\lambda$ for each $k$ as a function of spin parameter.
For g modes ($k\ge 0$), in the limit of slow rotation, 
\begin{equation}
\lambda \approx \ell(\ell+1) \quad {\rm if} \quad s\ll 1, 
\end{equation}
with $\ell = |m| + k$.

Except for prograde sectoral g modes ($k=0$ \& $m<0$)\footnote{In this paper we adopt the convention that prograde modes correspond to negative azimuthal order $m$ [cf. equation~(\ref{eq:func})]}, $\lambda$ for tesseral and retrograde g modes becomes very large as the spin parameter increases.
For a large $\lambda$, pulsation amplitude is strongly confined to the equatorial region as can be seen from surface patterns ($\Theta_k^m\cos m\phi$) shown at some places in Fig.~\ref{fig:lambda}.  Those retrograde and prograde tesseral modes should have small visibility because of cancellation. 

On the other hand, for prograde sectoral g modes we have
\begin{equation}
\lambda(s;k=0) \approx m^2 \quad {\rm if} \quad s\gg 1;
\end{equation}
i.e., $\lambda$ decreases as $s$ increases. Therefore, the amplitude distribution of a sectoral prograde g mode is less affected by rotation so that the visibility should be much higher than tesseral or retrograde g modes in a  rapidly rotating star. The prediction is consistent with the fact that prograde sectoral g modes are predominantly detected in rapidly rotating $\gamma$ Dor stars \citep{vanr16,oua17,zwi17} and SPB (Slowly Pulsating B) stars \citep{pap17}.   

Lines of $\lambda$ for r modes ($k < 0$), appear in the lower left (retrograde) side of Fig.~\ref{fig:lambda}. 
R modes are present if $\lambda > 0$; i.e., only if $s > (m + |k|)(m+|k|-1)/m$. In other words, the frequencies of r modes are bounded, in the co-rotating frame, as 
\begin{equation}
\omega({\rm r~mode}) < {2m\Omega\over (m+|k|)(m+|k|-1)} \le \Omega,
\label{eq:omr}
\end{equation}
where the last inequality comes from $k\le-1$ and $m\ge1$ (equality occurs in the case of $k=-1$ with $m=1$).

For the cases of $k\le -2$, $\lambda$ remains small even for a large value of $s$ as 
\begin{equation}
\lambda(s;k\le-2) \approx {m^2\over(2|k|-1)^2}  \quad {\rm if} \quad s \gg 1
\end{equation}
\citep{tow03a}. Amplitude of these modes is not confined to the equatorial zone, while $\lambda$ for $k=-1$ modes becomes very large as for retrograde g modes and the amplitude is confined strongly to the equatorial zone.

\begin{figure}
  \includegraphics[width=0.49\columnwidth]{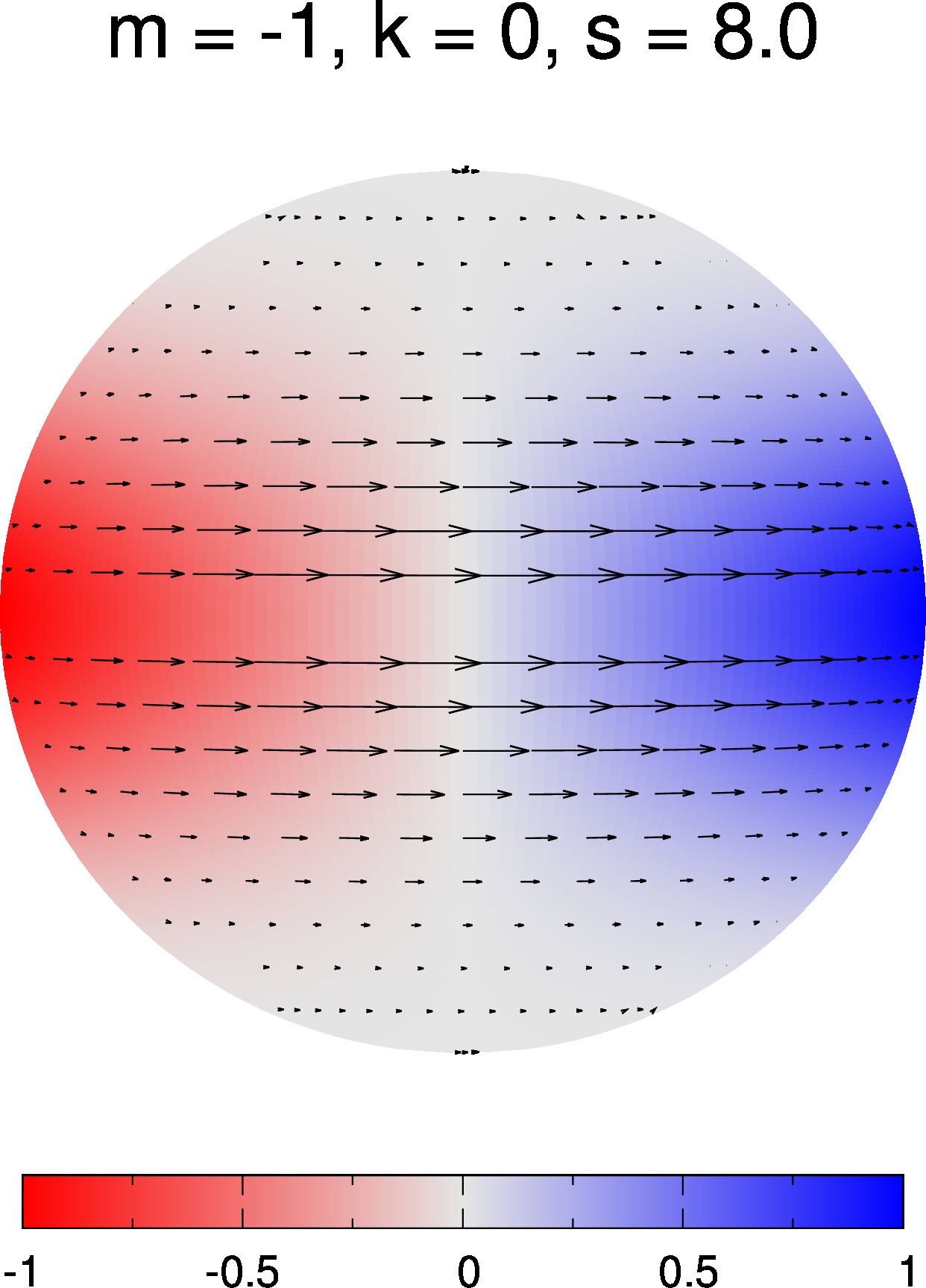} %{/Users/saio/rotationeffect/trad_appro/pattern/m-1k0_spn8p0_hdisp_10cm.jpg}
  \includegraphics[width=0.49\columnwidth]{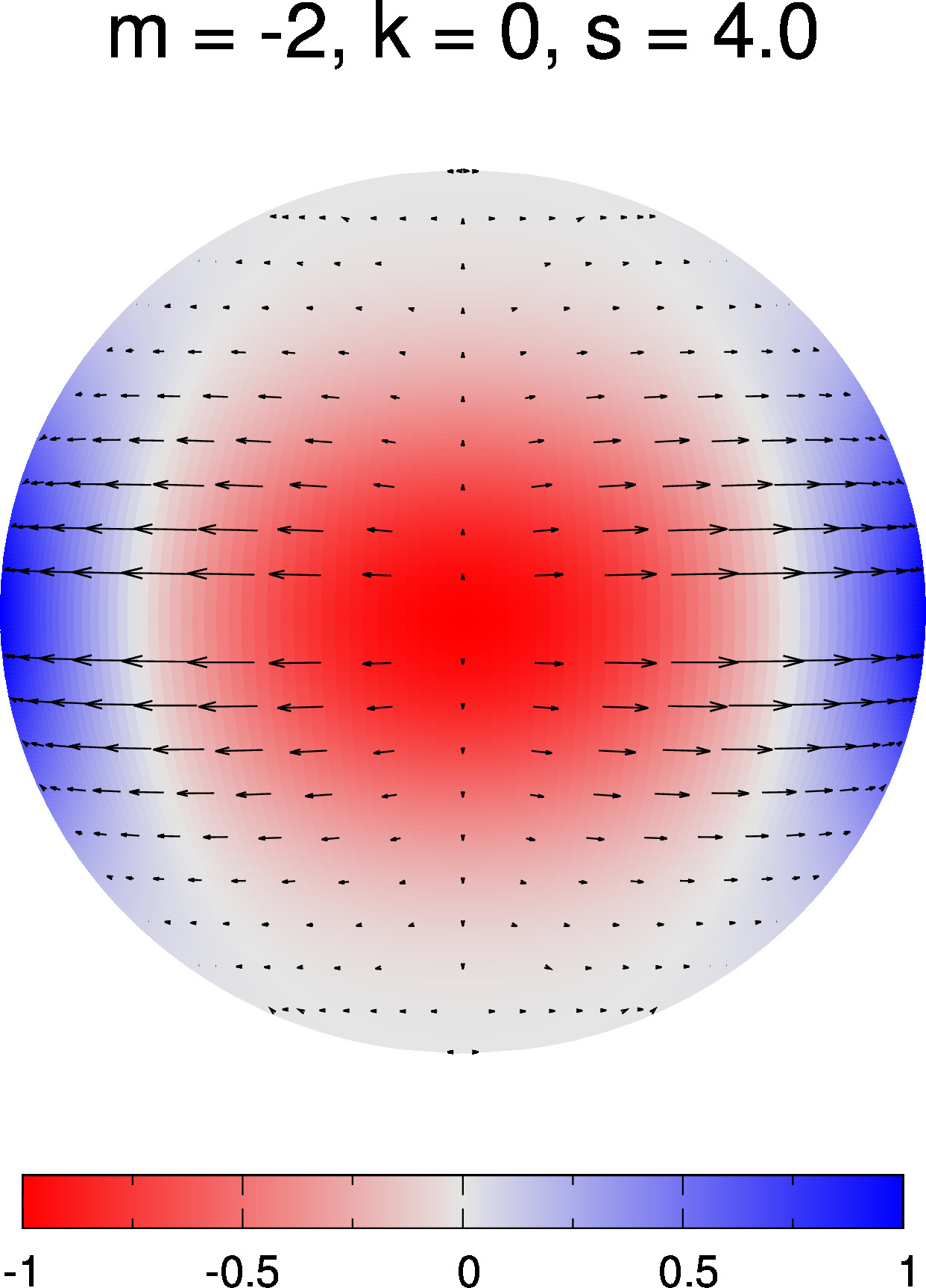} %{/Users/saio/rotationeffect/trad_appro/pattern/m-2k0_spn4p0_hdisp_10cm.jpg}
  
 \vspace{0.5cm}
  \includegraphics[width=0.49\columnwidth]{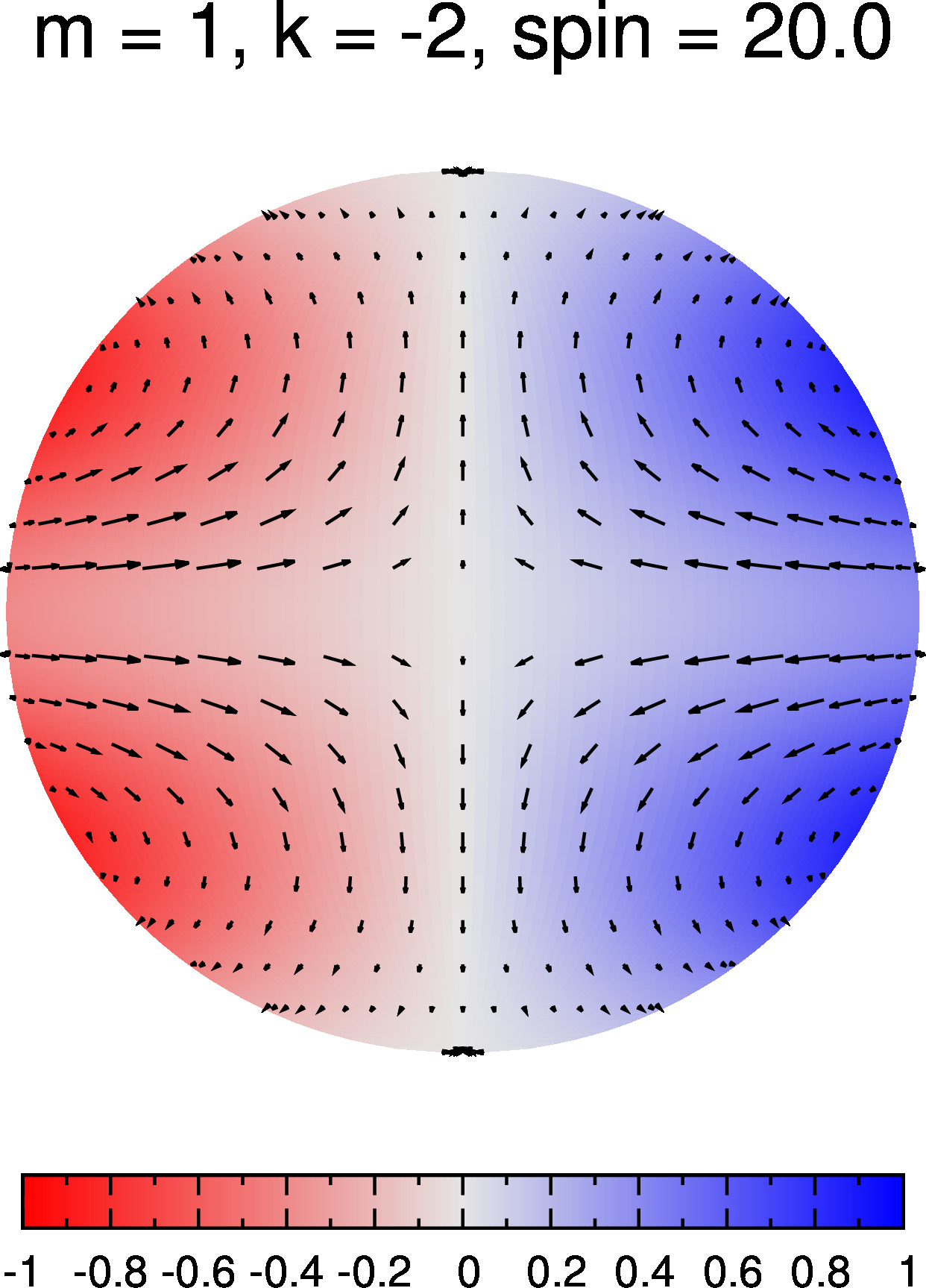} %{/Users/saio/rotationeffect/trad_appro/pattern/m1k-2_spn20_vel_10cm.jpg}
  \includegraphics[width=0.49\columnwidth]{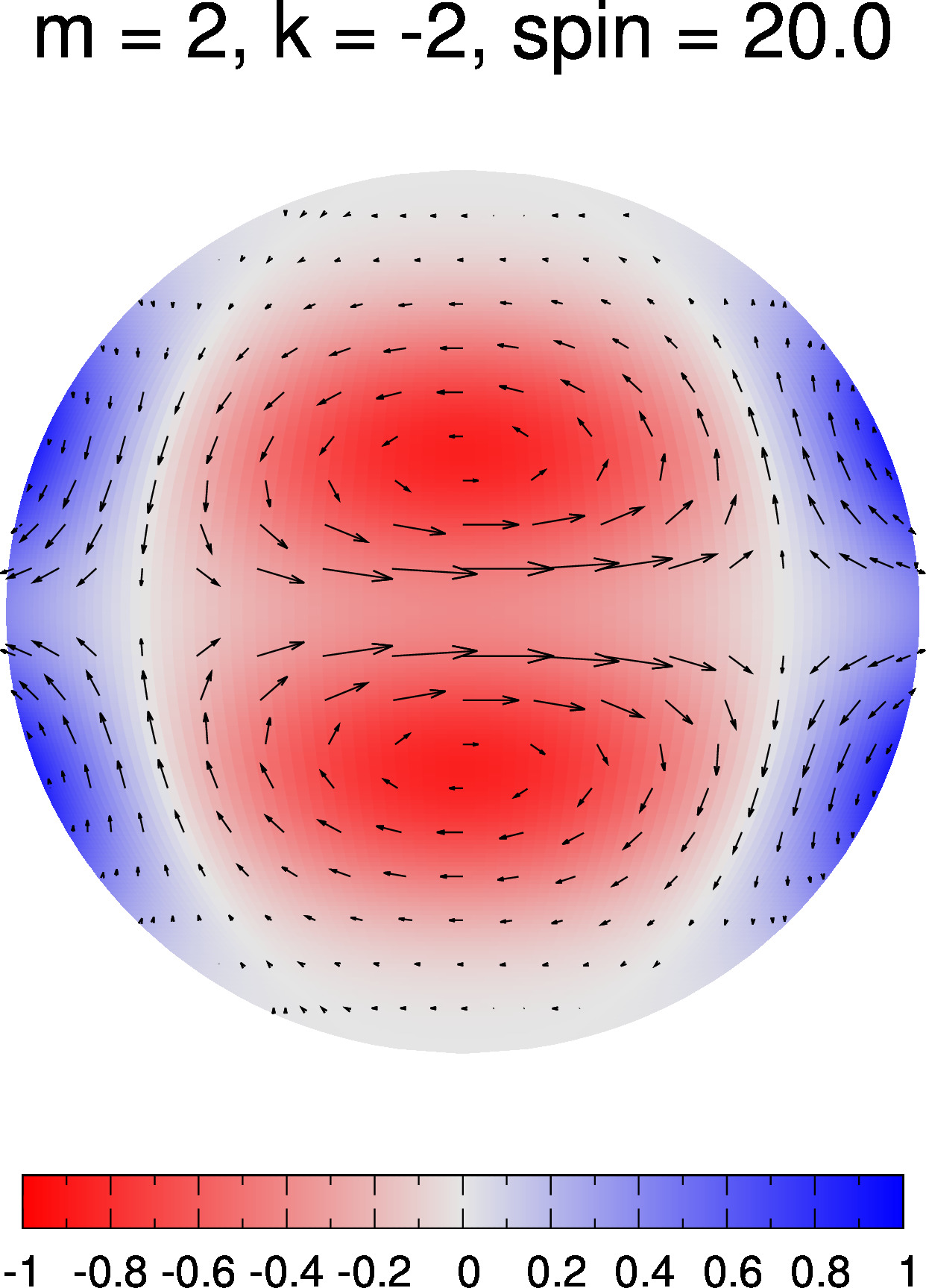} %{/Users/saio/rotationeffect/trad_appro/pattern/m2k-2_spin20_vel_10cm.jpg} 
  \caption{Upper two panels show displacement vectors and temperature variations on the surface for prograde sectoral g modes of $m=-1$ (left) and $-2$ (right). Lower two panels show horizontal velocity vectors and temperature variations for even r modes of $m=1$ (left) and $2$ (right).}
  \label{fig:pattern}
\end{figure}

Figure~\ref{fig:pattern} compare amplitude distributions of prograde sectoral g modes of $m= -1$ and $-2$ (upper panels) and even r modes of $m=1$ and $2$ with $k=-2$ (lower panels).
Arrows show horizontal displacements for g modes and flow vectors for r modes. 
For g modes horizontal compression (expansion) causes positive (negative) temperature perturbation, while for r modes positive temperature perturbation is generated by the Coriolis force with clockwise (counter-clockwise) motion in the northern (southern) hemisphere.  (Vorticity anti-symmetric with respect to the equator generates symmetric  temperature perturbation.)

Because the set of equations~(\ref{eq:tar1})(\ref{eq:tar2}) is the same as that of non-rotating case if $\lambda$ is replaced with $\ell(\ell+1)$, we can use asymptotic relations of g modes derived for a non-rotating star by replacing $\ell(\ell+1)$ with $\lambda$. 
Then, the cyclic frequency  of a high order ($n_{\rm g}\gg 1$) r or g mode in the co-rotating frame is given as
\begin{equation}
	\nu_{\rm co}({\rm g, r})\approx{\sqrt{\lambda}\over2\pi^2n_{\rm g}}\int_{r_1}^{r_2} \frac{N}{r}dr \equiv {\sqrt{\lambda}\over n_{\rm g}}\nu_0
\end{equation}
\citep{lee87,bou13}. It is noteworthy that the equation for the r-mode frequencies in the co-rotating frame is formally the same as that for g modes. Difference is brought only by different values of $\lambda$.  

\begin{figure}
 \includegraphics[width=\columnwidth]{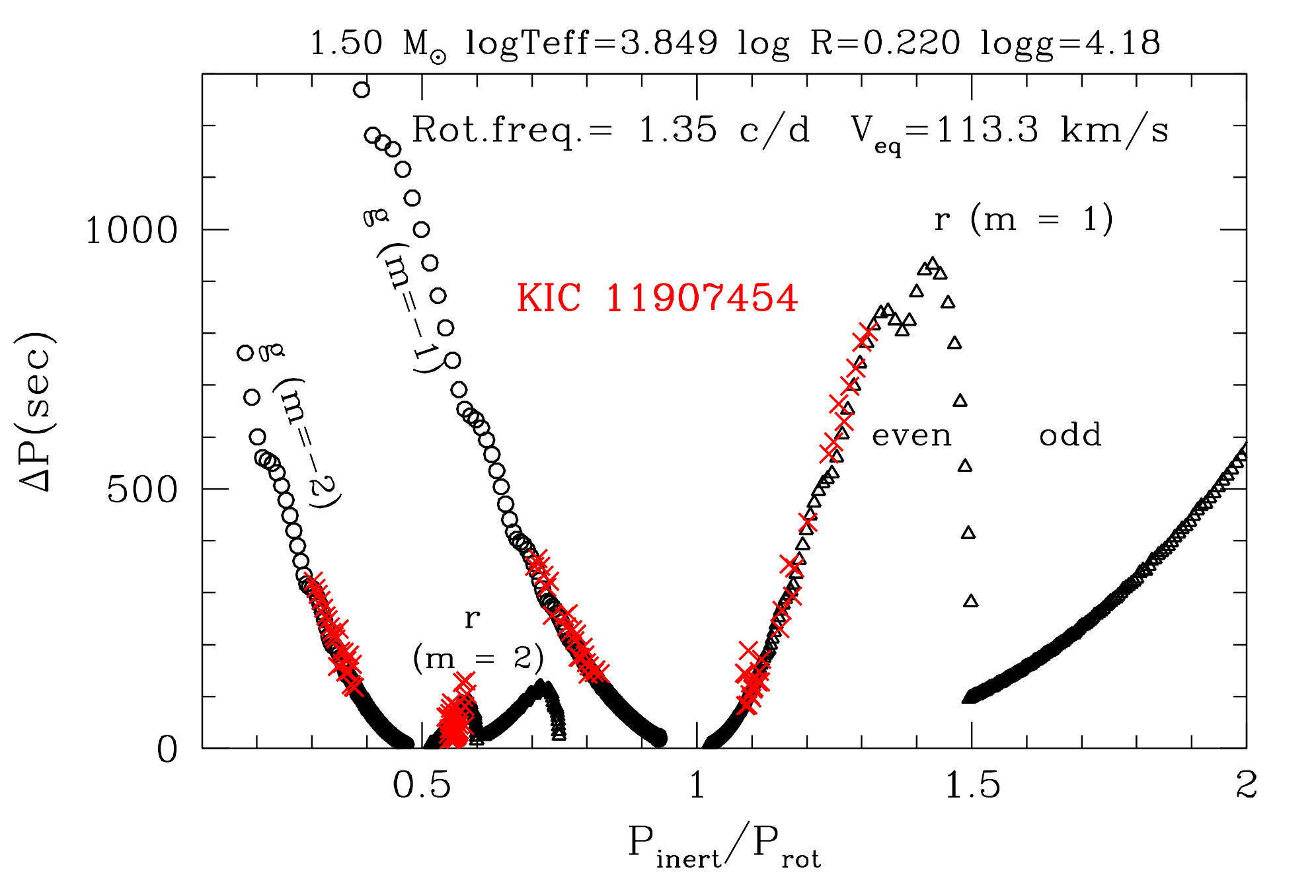}  %{p_dp_m1p50_1p35cd_k11907454.pdf}
\caption{Period spacings for r modes and prograde sectoral g modes as a function of period normalized by the rotation period. Red crosses are period spacings of the $\gamma$ Dor star  KIC~11907454 obtained from analysing the Kepler data. Periods (in observer's frame) of r modes of azimuthal order $m$ are longer than $P_{\rm rot}/m$, while periods of prograde sectoral g modes are shorter than $P_{\rm rot}/|m|$.   }
\label{fig:p_dp}
\end{figure}

In the inertial frame, the difference is more apparent; we have
\begin{equation}
  \nu_{\rm inert}({\rm g, r}; m) = \left|{\sqrt{\lambda}\over n_{\rm g}}\nu_0 - mf_{\rm rot}\right|,
\label{eq:nugr}
\end{equation}
where $f_{\rm rot}\equiv\Omega/(2\pi)$, cyclic frequency of rotation.
For prograde ($m<0$) sectoral ($k=0$) high-order ($n_g\gg1$) g modes with $s\gg1$, $\lambda \approx m^2$ (Fig.~\ref{fig:lambda}), so that we have 
\begin{equation} 
\nu_{\rm inert}({\rm g}; k=0, m<0) \approx |m|\left({\nu_0\over n_{\rm g}}+f_{\rm rot}\right).
\label{eq:mg}
\end{equation}
This equation explains why g-mode frequencies in rapidly rotating $\gamma$ Dor stars tend to  make groups (each group is associated with a value of $|m|$) \citep{mon10,sai18b}. In addition, from equation~(\ref{eq:mg}), we can derive a relation that period spacings of g modes should decrease with period as Fig.~\ref{fig:p_dp} shows.

\begin{figure}
\includegraphics[width=\columnwidth]{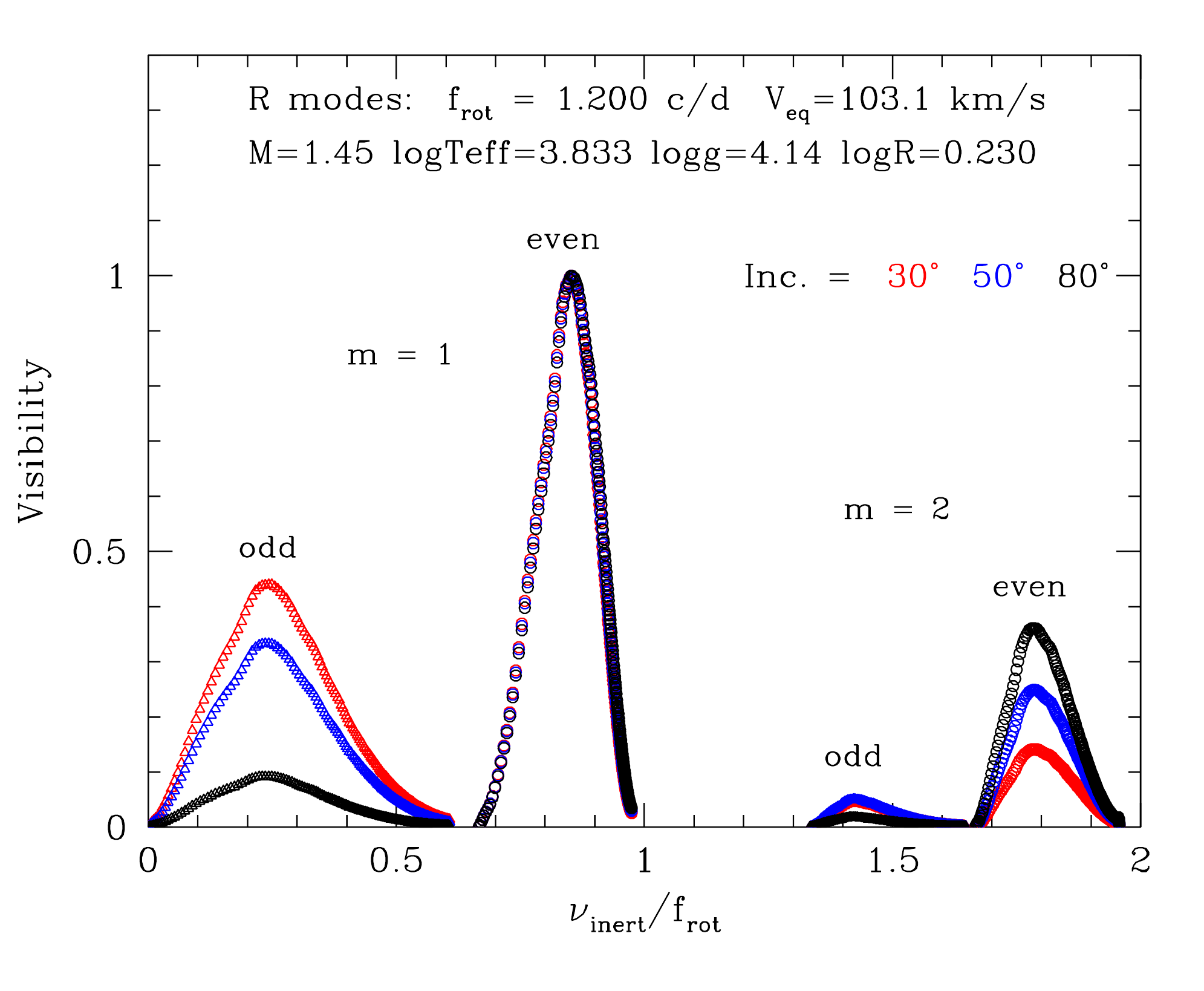}  %{rmode_vis_ffrot_inc853.pdf}
\caption{Visibility of r modes expected if kinetic energy is equally distributed. Circles and triangles are even and odd modes, respectively. Colour-coded three cases of inclination are normalized at the maximum of $m=1$ even modes. Horizontal axis measures frequencies in the inertial frame relative to the rotation frequency.}
\label{fig:vis}
\end{figure}

Equation~(\ref{eq:omr}) indicates r-mode frequencies in the co-rotating frame to be smaller than the rotation frequency; i.e., $\nu_{\rm co}({\rm r}) < f_{\rm rot}$. Therefore, equation~(\ref{eq:nugr}) for r modes should be written as
\begin{equation}
  \nu_{\rm inert}({\rm r};m) = mf_{\rm rot}-\nu_{\rm co}({\rm r})= mf_{\rm rot} - {\sqrt{\lambda}\over n_{\rm g}}\nu_0.
\label{eq:rnu}
\end{equation} 
This means that frequencies of r mode of azimuthal degree $m$ are bounded as
\begin{equation}
(m-1)f_{\rm rot} < \nu_{\rm inert}({\rm r};m)< mf_{\rm rot}
\label{eq:range}
\end{equation}
(cf. Figs.~\ref{fig:p_dp},\ref{fig:vis}).
As seen in Fig.~\ref{fig:p_dp}, period spacings of r modes generally increase with period (except for the region corresponding to a rapid change of $\lambda$ close to zero).
The dependence is opposite to the case of prograde sectoral g modes. The difference is caused by the negative sign for the last term in equation~(\ref{eq:rnu}) \citep[see][for details]{sai18b}.  

Observed frequencies of r modes lie much narrower ranges than equation~(\ref{eq:range}) \citep{vanr16,sai18}. We need to take into account the visibility of r modes.  
The pattern of temperature variation on the surface and hence the visibility of a mode depends on $s$, $m$, and $k$. 
The visibility of each r mode is calculated by integrating $\Theta_k^m(\theta;s)$ over the projected visible surface assuming a standard limb-darkening coefficient of $\mu=0.6$.
The visibility is divided by the square-root of the kinetic energy of the mode to have a non-dimensional quantity \citep[see][for details]{sai18}.
Thus, the visibility distribution among r modes corresponds to the amplitude distribution expected if energy is equally distributed among the r modes. 

Figure~\ref{fig:vis} shows visibility distributions for selected inclination angles for a main-sequence model with a moderate rotation frequency.
Generally, r-mode frequency distribution is very dense, and $m=1$ even modes are most visible, whose frequencies are close to but slightly less than the rotation frequency.
As expected, odd modes are less visible for higher inclination angles, and the visibilities for $m=2$ modes are smaller than those of $m=1$.

\section{Examples}
\citet{sai18} discussed the presence of r modes in $\gamma$ Dor stars, early-type stars with "hump\&spike" feature \citep[called RotD in e.g.,][]{bal13,bal15,bal17}, Be stars, and eclipsing binary stars  (including  heartbeat stars).  
In the following of this paper, we discuss the presence of r modes in the enigmatic F-type star KIC~8462852, early-type chemically peculiar Am and HgMn stars, and an accreting white dwarf in the dwarf nova GW Lib. 

\subsection{KIC~8462852 (Boyajian's star)} 
\begin{figure}
   \includegraphics[width=\columnwidth]{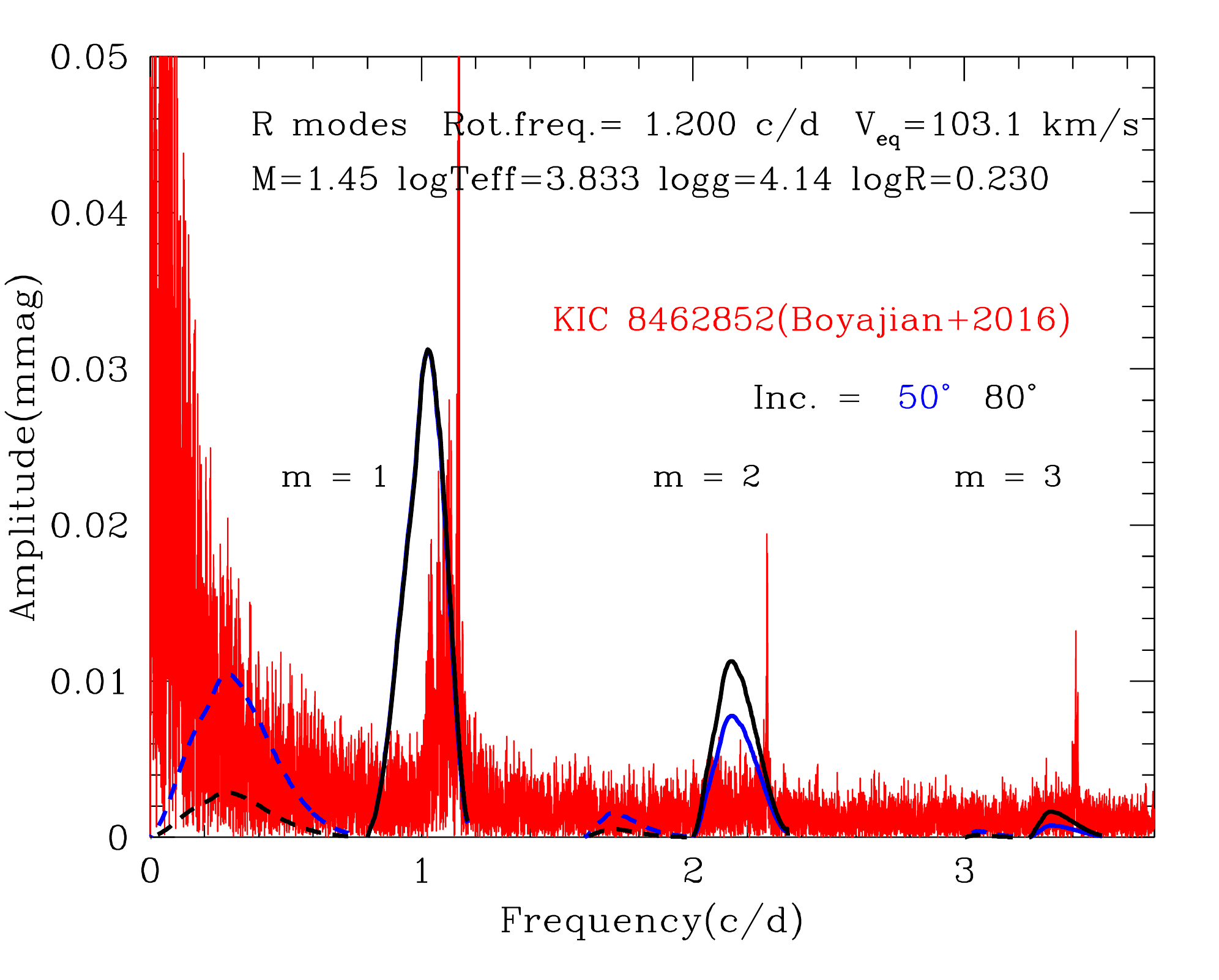}  %{boya_rmodes_m1p45_20_1p20cd.pdf}
   \caption{Fourier amplitude spectrum (red lines) of Boyajian's star (KIC~8462852) is compared with expected visibility distributions of r modes of $m=1, 2$, and $3$ for a main-sequence model of $1.45~M_\odot$, in which rotations frequency is assumed to be $1.20$~d$^{-1}$. Solid and dashed lines are for even and odd modes, respectively. Blue and black lines are for inclinations of $50^\circ$ and $80^\circ$. Visibility distribution is arbitrarily normalized at the maximum of $m=1$ even mode, for which blue and black lines are almost perfectly superposed. }
   \label{fig:boya}
\end{figure}

KIC~8462852 is an enigmatic star found by \citet{boy16} showing `irregularly shaped aperiodic dips in flux', but the star itself is a normal main-sequence F3~V star.
\citet{boy16} also found three equally-spaced "hump\&spiks" (a dense agglomeration of small-amplitude peaks to the left of a sharp peak) in a Fourier amplitude spectrum of  KIC~8462852.
These three dense groups can be identified as even r modes of $m=1, 2$, and $3$.
The visibility distributions of those r modes in a main-sequence model of $1.45~M_\odot$ are superposed on the observational amplitude spectrum of KIC~8462852 in Fig.~\ref{fig:boya}. A rotation frequency of $1.2$~d$^{-1}$ is assumed to get a best fit.  
The adopted values of $T_{\rm eff}$ and $\log g$ are consistent with the spectroscopic values listed in \citet{boy16}. Assumed rotation frequency $1.2$~d$^{-1}$ is very close to but  slightly larger than the frequency $1.14$~d$^{-1}$ at the main sharp peak. 
\citet{boy16} obtained a projected rotation velocity of $V\sin i = 84\pm 4$~km~s$^{-1}$ for KIC~8462852.  Combining the $V\sin i$ with the equatorial velocity $103$~km~s$^{-1}$ of the model gives an inclination angle of $55^\circ \pm 4^\circ$. 

Probably r modes would not give a solution for the enigmatic drippings in flux of KIC~8462852, but the presence of r modes in this star itself is interesting.

\subsection{Am star: KIC~9117875}
\begin{figure}
  \includegraphics[width=\columnwidth]{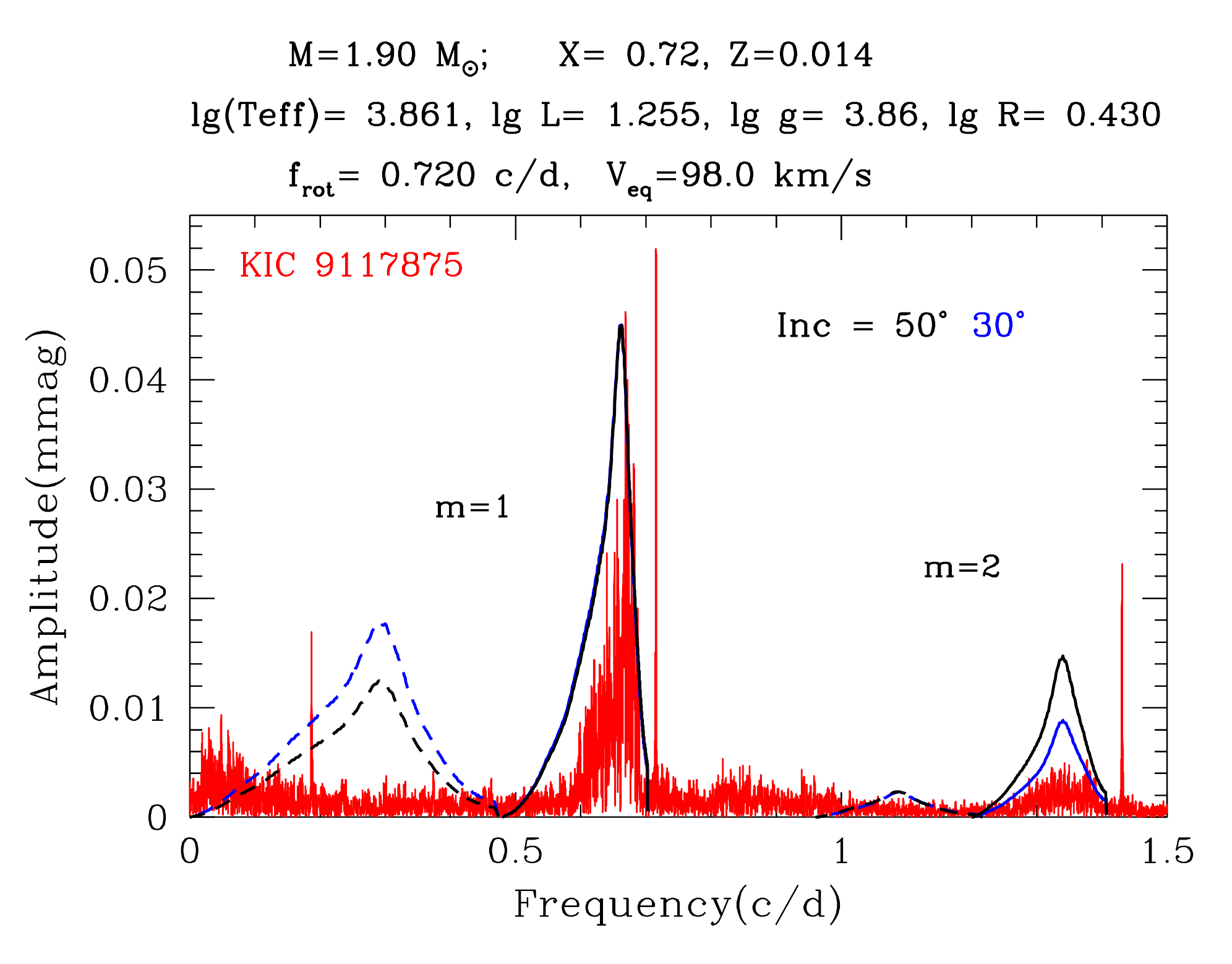}  %{rmodes_m1p9x72z014_18_0p72cd_k9117875_Am_v2.pdf}
  \caption{Amplitude spectrum of the Am star KIC~9117875 (= HD~190195) is compared with visibility distribution of r modes calculated for a $1.9$-$M_\odot$ main-sequence model with a rotation frequency of $0.72$~d$^{-1}$. The visibility distribution is normalized arbitrarily.}
  \label{fig:am}
\end{figure}

Chemically peculiar Am (CP1) stars are A-F type stars having spectra with enhanced metallic lines, which are caused by radiative levitation and gravitational settling.
It is known that majority of Am stars do not pulsate even if they are located in the $\delta$ Sct instability strip \citep[e.g.][]{sma11,mur15}. 
The reason is thought as the drainage of helium from the He II ionization zone due to gravitational settling, which weakens the kappa-mechanism excitation for pulsation \citep{bag72}.

Using Kepler light curves, \citet{mur15} identified not-pulsating (in p modes) Am stars (and normal stars) within the $\delta$ Sct instability strip. 
Inspecting Fourier amplitude spectra of those stars, we find signatures of r modes (hump\&spike) in many cases.
As an example, Fig.~\ref{fig:am} shows a Fourier amplitude spectrum of KIC~9117875 \citep[= HD~190165; No.42 in][]{mur15}. The amplitude spectrum is compared with the visibility distribution of r modes predicted for a main-sequence model of $1.9~M_\odot$ with a rotation frequency of $0.72$~d$^{-1}$.
The amplitude spectrum of KIC~9117875 shows spikes that are probably caused by a spot (or two spots). The frequency at the main spike is $0.716$~d$^{-1}$, which should indicate the rotation frequency at a spot of the star. This supports the assumed rotation frequency, $0.72$~d$^{-1}$, for the model. Corresponding equatorial velocity $98$~km~s$^{-1}$ is consistent with  a spectroscopic value of $V\sin i= 61 \pm3$~km~s$^{-1}$ obtained by \citet{nie15}. For the amplitude ratio between $m=1$ and $2$, a smaller inclination such as  $i \sim 30^\circ$ is favoured.

We note in passing that most chemically-normal non-pulsators lie in the $\delta$ Sct instability strip that are identified by \citet{mur15} show signatures of r modes in their Fourier amplitude spectra. 

\subsection{HgMn star; KIC~6128830}
\begin{figure}
	\centering
	\includegraphics[width=\columnwidth]{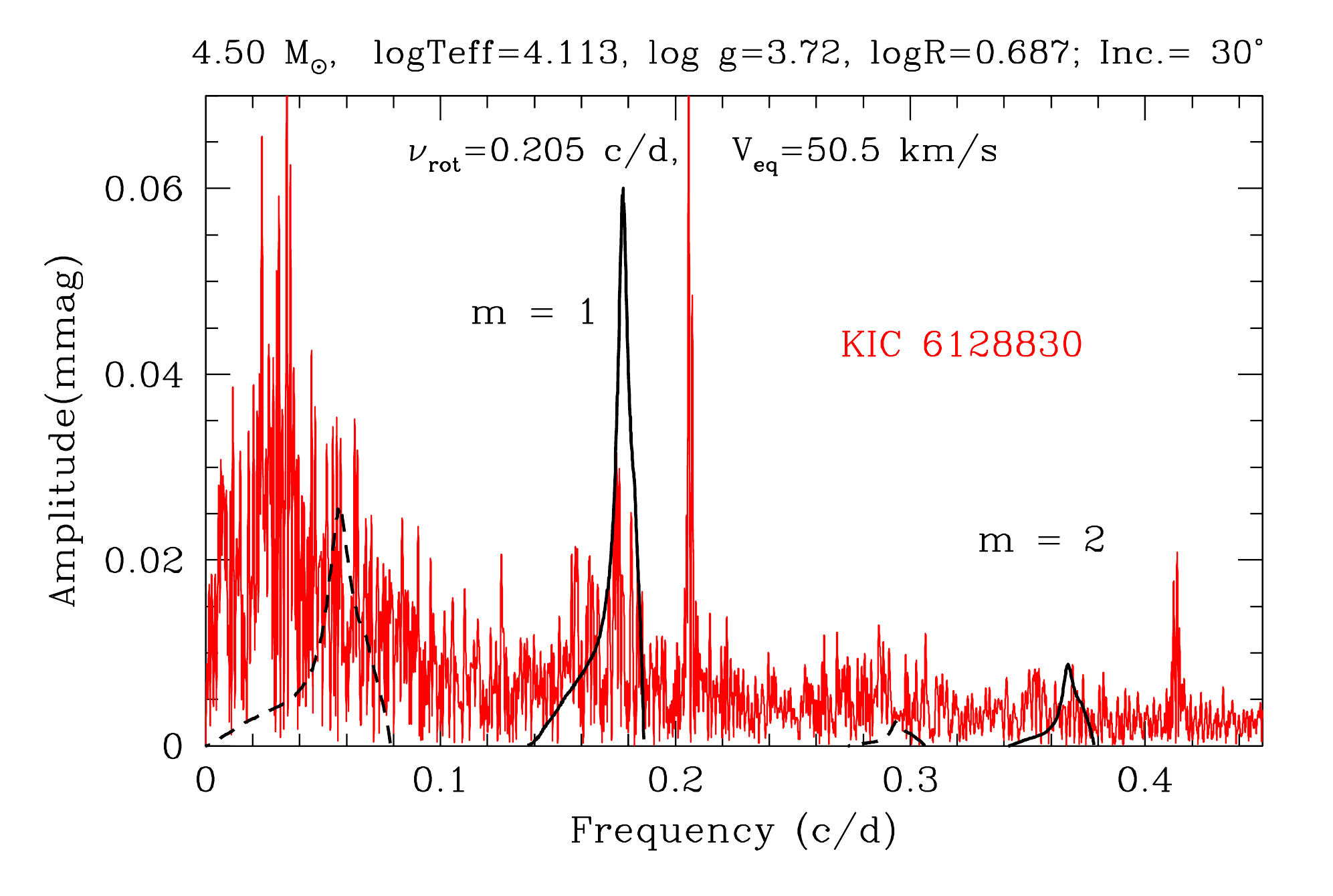}  %{rmodes_m4p5_26_0p205cd_k6128830_inc30.pdf}
	\caption{The amplitude spectrum of HgMn star KIC~6128830 is compared with visibility distribution of r modes predicted for a main-sequence model of $4.5~M_\odot$ with a rotation frequency of $0.205$~d$^{-1}$. Solid and dashed lines are even and odd r modes, respectively. The visibility distribution is normalized arbitrarily such that the maximum visibility corresponds to an amplitude of $0.06$~mmag. }
	\label{fig:HgMn}
\end{figure}
Chemically peculiar HgMn (CP3) stars are slow rotators, and have atmospheres with Hg and Mn being strongly enhanced, while He being depleted. The effective temperature lies between 10 kK to 16 kK.  
\citet{hum18} confirmed that KIC~6128830 is a HgMn star having Hg and Mn enhanced by factors of $\sim\!\!10^5$ and $\sim\!\!300$, respectively, with a slow rotation velocity of $V\sin i = 25 \pm 2$~km~s$^{-1}$.
They also obtained spectroscopic parameters of $\log T_{\rm eff}= 4.11$ and $\log g = 3.5$.
Analysing the Kepler light curve of KIC~6128830, \citet{hum18} found nearly sinusoidal light variations associated with a frequency of $0.2065$~d$^{-1}$ with harmonics, and concluded it is caused by a spot (or spots) of chemical composition.

In addition to the main sharp peaks, \citet{hum18} found small amplitude residual peaks to the left  of the main frequency and of the second harmonic.
Those residual peaks can be identified as r modes of $m=1$ and $2$.  Fig.~\ref{fig:HgMn} shows visibility distributions of r modes for a model with a rotation frequency of $0.205$~d$^{-1}$.
Frequency ranges of r modes are roughly consistent with the small amplitude peaks of KIC~6128830, although visibility distributions for even modes are too steep near the maximum. 
Although theoretical amplitude distribution is not very sensitive to the inclination angle, an inclination of $30^\circ$ is chosen for the equatorial velocity to be consistent with the spectroscopic $V\sin i$. 
 
It is interesting that r modes are present in such a slow rotator as KIC~6128830.
Because the star has a strong feature indicating the presence of a spot (or spots), r modes may be excited mechanically by `meandered' rotation flow around a spot.
R modes may also be excited thermally by the kappa-mechanism at the Fe opacity peak at $T \sim 1.5\times 10^5$~K \citep{tow05,sav05,lee06,dzi07} because KIC~6128830 is located around the edge of the SPB (Slowly Pulsating B stars) instability region in the $\log T_{\rm eff}-\log g$ diagram.   

\subsection{Accreting white dwarf in the dwarf nova GW Lib}

\begin{figure}[t!]
   \includegraphics[width=\columnwidth]{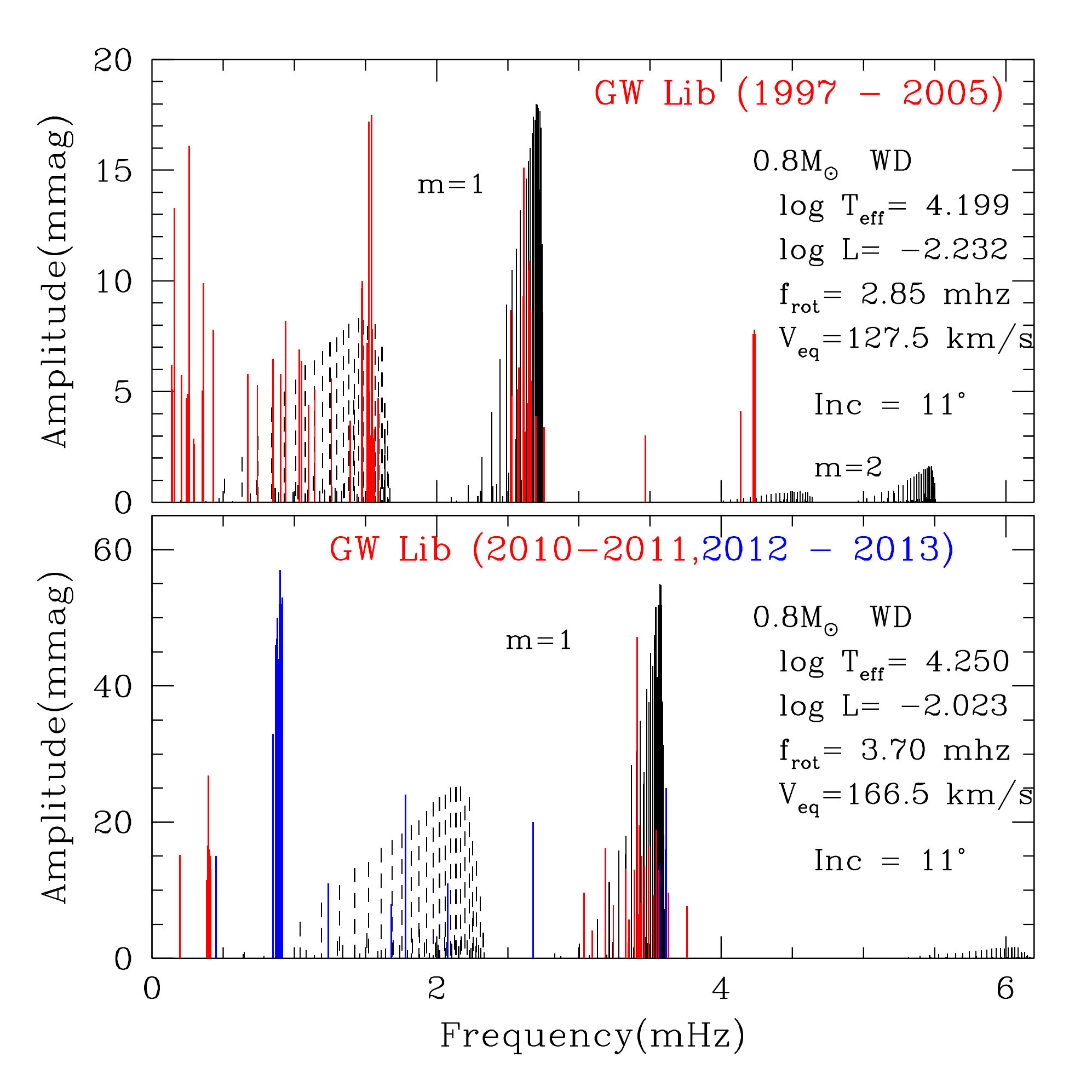}  %{freq_amp_rmodes_m0p8WD_GWLib_v6.pdf}
   \caption{Frequency-amplitude diagrams for the dwarf nova GW Lib before (upper panel) and after the outburst in 2007. Superposed (black lines) are expected visibility distributions of r modes of $m = 1$ and $2$ in $0.8$-$M_\odot$ white dwarf models, where solid and dashed lines are even and odd modes, respectively. Rotation frequencies of 2.85~mHz before the outburst (upper panel) and $3.70$~mHz after the outburst (lower panel) are assumed.  The visibilities are normalized arbitrarily at the maxima of the $m=1$ even modes (at $\sim2.7$~d$^{-1}$ in the upper panel and $\sim3.6$~d$^{-1}$ in the lower panel.}
   \label{fig:accwd}
\end{figure}

Pulsations in an accreting white dwarf were first discovered by \citet{war98} in the quiescent phase of the dwarf nova GW Lib.
Since then, pulsations have been detected in quiescent phases of nearly twenty dwarf novae \citep[see e.g.][for a review]{szk13}.
They belong to WZ Sge type dwarf novae having very low accretion rates of $\sim\!\!10^{-11}M_\odot$yr$^{-1}$ in the quiescent phase, and have very large outbursts in very long intervals (a few decades).
Because of the very low accretion rate in the quiescent phase, a large fraction of the flux from a system is originated from the white dwarf, so that nonradial pulsations of the white dwarf can be detected.

GW Lib is a well studied representative case of pulsating accreting white dwarfs. In addition, GW Lib is one of the few cases where nonradial pulsations were observed during  quiescent phases before and after an outburst.  
GW Lib had a very large (9~magnitude) outburst in 2007 after 24 years of the discovery in 1983.

The upper panel of Fig.~\ref{fig:accwd} shows pulsation frequencies/amplitudes detected before the 2007 outburst adopted from \citet{vanz04} (1997--2001), \citet{szk02} (2002 Jan), and \citet{cop09}(2005 May), while the lower panel shows frequencies/amplitudes obtained after the outburst adopted from \citet{szk12}(2010 Mar-2011 Aug), \citet{cho16}(2012 May), and \citet{szk15} (2012Jun,2013Mar).
All frequencies/amplitudes reported in the literature are shown in this figure (coloured lines). Each observing run detects only a few frequencies which are similar but slightly different from the frequencies previously obtained; this is probably due to poor resolution because of a short baseline of each run, and beating among densely distributed frequencies, which is common for r-mode oscillations. 

Black lines show theoretical prediction for frequencies and visibilities of r-mode oscillations in $0.8$-$M_\odot$ white dwarf models.  
(cf. \citet{vanspa10apj} obtained $0.84~M_\odot$.)
Slightly higher effective temperature for the model after the outburst is adopted taking into account the accretion heating \citep{szk12}, although the property of r modes is not sensitive to $T_{\rm eff}$. 

In contrast to the main-sequence models, visibility of r modes in white dwarfs does not vary smoothly among adjacent modes, because the degree of wave reflection at a steep H/He transition is sensitive to oscillation frequency. Visibility is much higher for r modes trapped in the hydrogen-rich envelope. 

To fit observed main frequency groups of GW~Lib, we have assumed rotation frequencies of $2.85$~mHz before the 2007 outburst and $3.70$~mHz after the outburst (rotation periods of $351$~s and $270$~s, respectively).
These rotation frequencies should represent rotation rates in the hydrogen-rich envelope.
Fig.~\ref{fig:accwd} shows that main frequency groups of GW~Lib are roughly consistent with $m=1$ r modes. (Low frequency features might come from the accretion disk).
The predicted equatorial rotation velocity after the outburst, $167$~km~s$^{-1}$ is consistent with the spectroscopic projected rotation velocity $V\sin i = 40$~km~s$^{-1}$ obtained by \citet{szk12}, because the inclination angle of GW~Lib is as low as $11^\circ$ \citep{vanspa10apj}.

Our r-mode models for the nonradial pulsations of GW~Lib indicate that the H-rich envelope of the white dwarf was considerably spun up by the accretion during the 2007 outburst.
Further monitoring of pulsations in GW~Lib would be very interesting.

Dwarf nova EQ~Lin is a similar system which had an outburst in 2006.
Frequencies of nonradial pulsations before and after the outburst were obtained by \citet{muk11,muk13,szk15}.
A preliminary r-mode fitting to those frequencies seems to indicate that an increase in rotation frequency from before to after the outburst is much smaller than the case of GW~Lib.    
Variety in the degree of spin-up is quite reasonable because the angular momentum given to the  white dwarf would be sensitive to the binary parameters. 
It is interesting and important to obtain spin-up information for other dwarf novae.

\section{Summary}
As we discussed above and in \citet{sai18}, r modes are present in various types of stars. They seem to be mainly excited mechanically by flows generated by spots, tidal forces, accretion, mass-loss, or g-mode oscillations. 
Mechanical excitation of r modes is expected to be easier than excitation of g modes. Since r mode motions are mainly toroidal, even non-organized horizontal motions may generate r modes due to the effect of Coriolis force, while to excite g modes, some organized spheroidal (with compression/expansion) motion must be generated.

If we detect r modes in a star, we can determine a rotation frequency of the star by fitting the frequency range (and period spacing if possible) with a model.
Obtaining rotation frequencies are very useful. For example, obtaining rotation frequencies of the member stars in a binary system would be useful to understand the process of orbital-rotational synchronization.  
Another example is accreting white dwarfs in dwarf novae, where each outburst caused by a disk instability pours much matter (and hence angular momentum) onto the white dwarf and spins up the star. By monitoring r mode frequencies, we can see the time-variation of the rotation rate in the H-rich envelope of the white dwarf. Needless to say, it is very important information to understand the evolution of angular momentum in the white dwarf.

\section*{Acknowledgments}
I am grateful to NASA and the Kepler team for their revolutionary data.

\bibliographystyle{phostproc}
\bibliography{ref_proc.bib}

\end{document}